# Intrusion Detection on Smartphones

Muhamed Halilovic
International Burch University
Faculty of Engineering and Information Technologies,
Department of Information Technologies,
Sarajevo, Bosnia and Herzegovina
mhalilovic@stu.ibu.edu.ba

Abdulhamit Subasi
International Burch University
Faculty of Engineering and Information Technologies,
Sarajevo, Bosnia and Herzegovina
asubasi@ibu.edu.ba

*Abstract* – **Smartphone technology is more and more becoming the predominant communication tool for people across the world. People use their smartphones to keep their contact data, to browse the internet, to exchange messages, to keep notes, carry their personal files and documents, etc. Users while browsing are also capable of shopping online, thus provoking a need to type their credit card numbers and security codes. As the smartphones are becoming widespread so do the security threats and vulnerabilities facing this technology. Recent news and articles indicate huge increase in malware and viruses for operating systems employed on smartphones (primarily Android and iOS). Major limitations of smartphone technology are its processing power and its scarce energy source since smartphones rely on battery usage. Since smartphones are devices which change their network location as the user moves between different places, intrusion detection systems for smartphone technology are most often classified as IDSs designed for mobile ad-hoc networks. The aim of this research is to give a brief overview of IDS technology, give an overview of major machine learning and pattern recognition algorithms used in IDS technologies, give an overview of security models of iOS and Android and propose a new host-based IDS model for smartphones and create proof-of-concept application for Android platform for the newly proposed model.**

## I. INTRODUCTION

Smartphone's operate in so called mobile ad-hoc networks or MANETS. Unlike to the infrastructure architecture, wireless ad hoc networks consist of group of mobile nodes or peers which are capable of communicating with each other without a physical access point. Topology of ad-hoc network is dynamic and is designed as a random graph. Other important aspects include the spacing of the nodes and their self-organizing abilities. Wireless medium is used to transmit data between different stations. Several different physical layers of architecture are defined to support the wireless medium access for mobile ad hoc networks to operate efficiently. MANETS are basically a system of mobile nodes where the communication between nodes is realized via wireless links. Due to mobility of nodes the topology of MANETs is often changing. Due to their inherent vulnerabilities, such as resource constrains, uncontrollable environment, and dynamic network topology, wireless ad hoc networks are subject to a variety of attacks [1].

Major problem with the intrusion detection systems that were developed for mobile-ware computer technologies is a problem of CPU and memory and their consumption, because smartphones and other mobile devices have limited energy resources. The other major problem is that most other protection systems such as antivirus need to continually update their virus signatures from the central repository and since updating of phone antivirus signatures is energy-expensive it is more likely that the attackers may try to use newer kinds of attacks to compromise smartphones.

Dikinson states that the antivirus companies take time to update their signature repositories, and when the new kind of malware is created and placed onto the network, hackers have substantial amount of time to perform their attacks before malware signatures are updated across different machines [2].

Other problem is that since there are many kinds of machines certain attacks can target specific kinds of operating systems and machines which might not be a priority for certain antivirus and security companies that serve updates to their customers. Often, smartphones fall into this category. Because of all these problems there is a need for more general approach to solving these issues. One of the major ways this can be achieved is by utilization of machine learning and design pattern algorithms, use of mathematical and statistical models which when applied are capable of detecting the malware which isn't discovered previously and which doesn't have to be within the repositories of the known malware.

The majority of work that deals with mobile IDS systems deals with host based IDS in which methods of anomaly or rule-based are utilized and used to extract and perform analysis of the features and then make decisions on the state of the device. The extraction is done locally and the analysis is done either locally or on a remote server.

Any smartphone IDS that would require high amount of computation on the device would lower the user experience, consume power resources, so the IDS that would truly be functional and protective for the user is extremely challenging to design and would have to be done on as lowest level OS as possible, carefully programming for evasion of memory leaks, etc.

## II. ANDROID SECURITY OVERVIEW

The following Android security overview is primarily based on official Android documentation.[1]

The main building blocks of Android device are Android hardware, Android operating system, and Android application runtime. The Android runs on all kinds of different hardware configurations. These include smartphones, tablets, set-up boxes.

The Android operating system is composed of the core which is build on top of Linux kernel. All the resources of the Android such as camera function, GPS-data, blue-tooth functions, telephony functions, network connections, are all accessed through the operating system.

The Android application runtime is created in Java programming language and is run in Dalvik virtual machine. However, many applications as well as Android core services and applications are native applications or integrate native libraries.

Dalvik as well as native applications all run inside of the same security environment which is contained within the applications sandbox. Every application has its own part of the file system where its private data including database and files can be written. The Android applications extend the Android core operating system and two main sources for applications are pre-installed applications and user-installed applications. Pre-installed applications include apps such as phone, email, calendar, browser and contacts. These are the key device capabilities and are available to the user as well as other applications. Pre-installed applications can also be developed by OEM-specific device.

The user installed applications are the third-party applications created by other developers. The Google Play Android web store contains hundreds of thousands of applications. Google also provides number of cloud-base services which can be accessed through the Android device.

Android provides the following key security features:
1) Robust Security at the OS level
2) Mandatory application sandbox for all applications.
3) Secure interpose communication
4) Application signing
5) Application defined and user granted permissions

At the level of the operating system the Android platform provides the security of the Linux kernel as well as secure process communication (IPC) facility which enables secure communication between applications running in different processes. Application send-box constraints even the native code. This ensures that if malicious application happens to get installed on the system it will be prevented from harming the system itself and harming other applications.

The Linux provides the Android with the following security features:
1) User based permission model
2) Process-based isolation
3) Extensible mechanism for secure IPC
4) The ability to remove potentially unsecure part of the kernel.

Each application which is installed on the Android OS behaves as a unique user. The Android assigns each user application unique ID and runs it as a separate user in a separate process.

The app signing enables the users to uniquely identify the author of the application and gives them ability to update their application without having to cope with complicated interfaces and permissions. When the developer creates the application he has to sign it. If there is an attempt to install an unsigned application the Google Play store will reject it as well as the packages installed on the Android device. This creates the mutual trust between the Google and the developer. The application can uniquely be identified and its developer and the developers can be accountable for the behaviour of their applications and developers can be assure that the application which is being sold or installed on users devices is really theirs.

Application signing is the first step in placing the application in application sandbox. The different applications run under different user IDs and the signed user certificate defines which user's id is associated with which application. The app signing makes sure that one application cannot access any other application except through well defined IPC.

Upon the installation of the APK file the package manager checks the APK for proper signature, and for the proper signing of the app included in the APK. If the certificate matches the key used to sign any other APK on the device the the APK has the option to specify in its manifest that it will share UID with the similarly signed APKs. The application can be signed by the third party or can be self-signed. The developers can generate a self-signing certificate without external assistance or permission (and this is the main difference with Apple's iOS). The applications do not have to be signed by the central authority (iOS applications do). The central authority verification is currently not mandatory for Android operating system.

---

[1] Official Android documentation:
http://source.android.com/tech/security/index.html

III. IOS SECURITY OVERVIEW

The iOS Security overview is primarily based on official Apple documentation.[2]

The iOS architecture at the highest level acts as an intermediary between and underlying hardware and the applications that appear on the screen. The apps that developer creates rarely talk directly to the underlying hardware. The interface between the application and the hardware are the system interfaces which protect the hardware from changes by the application.

This also makes it possible to easily create apps which work consistently on devices with different hardware capabilities.

The implementation of iOS technologies can be viewed as a set of layers which are shown in Figure 13, layers of iOS. The lowest layer is the system and it contains the most fundamental services and technologies upon which applications rely.

More sophisticated services and technologies are placed on higher layers. Even though the lower layer frameworks are available to the developers, developers should mostly stick with the higher level frameworks.

Besides having a built in security framework iOS also gives an option to use an explicit security framework, security dot framework that can be utilize to guarantee the security of the data application manages. This framework provides interfaces for managing certificates, public and private keys and trust policies. It includes the generation of cryptographically secured pseudorandom numbers. It also supports the storage of certificates and cryptographic keys in the keychain which is as we previously mentioned secure repository for user data. The common crypto library provides additional support for symmetric encryption HMAC, and digests. The digests feature provides functions that are essentially compatible with those in the OpenSSL library, which is not available in iOS.

J. Zdziarski states that iOS operating system incorporates four layers of security for its users and their data protection and these include [3] :

1) Device Security – methods/techniques to protect the device from unauthorized usage;
2) Data Security – methods/techniques to protect data on the device, even in case when device is stolen;
3) Network Security – encryption techniques used to encrypt data that is being sent over a network
4) Applications security – methods/techniques used to secure operating system and isolate individual applications while they are running.

iOS puts each application in a sandbox at the install time and this includes its preference and its data. A sandbox is a set of fine-grained controls which limit the apps access to resources of other applications such as files, preferences, network resources, hardware and etc. Upon its installation an application receives its own sandbox directory which serves as a home for the app and its data.

The main function of the sandbox model is to prevent malicious application from damaging other applications and their data. This means that application itself which does contain code which can result in a security breach may result in a damage of only that specific application. The sandbox prevents the hijacked app from affecting other applications and other parts of the system. This approach considers an environment in which the code is understood not to be trusted by default and is because of that isolated from processes and resources available to the operating systems. The apples Sandbox limits the number of CPU cycles an application can use as well as the amount of memory application can use. It also restricts apps from accessing files from outside its dedicated home directory.

Classes for interfacing with the camera, GPS and other resources on the devices are provided by the Apple, and restrictions are placed on applications from accessing many of the components directly. Besides this the binary code which is running on the device must be signed by Apple's central authority and this was incorporated by the Apple into its security model from the beginning.

Apple signs the application and this is the only way an application is permitted to run under the iOS. The reason for this is to make sure the applications haven't been modified from their original binary. The Apple also tests the integrity of an application in order to check whether there was unsigned code injected into the application. Another part of application security incorporated into Apples devices is the encrypted key-chain which provides central facility for storage and retrieval of network passwords, networking credentials and other information. The Apples security framework facilitates low level functionality for reading and writing the data to and from the keychain and performing encryption and decryption. The data within the key-chain is logically zoned so that the application cannot access the encrypted data stored by different applications[3]. Apple's Common Crypto architecture provides cryptographic APIs to the developers who would like to make use of encryption in their applications.

IV. IDS ON SMARTPHONES – RELATED WORK

---

[2] Official Apple documentation: http://developer.apple.com/library/ios/#documentation

Samfat and Molva propose architecture for mobile networks - IDAMN and they use anomaly detection methods as well as rule-based methods. There are three levels of detection; Location based detection (user located at two different locations at the same time), traffic anomaly detection (extremely low or extremely high levels network traffic), and the detection of anomalous behaviour of specific mobile phone user[4].

Artificial neural networks were utilized in order to detect anomalous behaviour of fraud in the usage of operator services, for example registration with the false identity and using the phone to hide tariff destinations. 16 distinct features representing standard mean deviation of the total duration and number of long and short term national and international calls were used for this kind of detection [5].

A collaborative proxy-based system for smartphones called Smartsiren was presented by Cheng and others[6]. By analyzing of communication activities of the smartphones they tried to detect abnormal behaviours on a single device as well as system-wide. Smartsiren is composed of backend proxy that communicates with light-weight agents on the protected devices. The information is collected and then sent to the proxy who then uses this to analyze it and if abnormal behaviour is detected the alert is sent.

Schmidt et all utilize a Symbian monitoring client for the smartphone based on Symbian OS, which collects and forwards collected features to anomaly detection server RADS[7]. This data is then processed in order to distinguish between normal and abnormal behaviour. The results show that top applications used by phone users modify and affect the monitored features in various ways. Intrusion detection systems that detect attacks which deplete the devices energy and basically reduce the quality of service to the user were analyzed by several papers [8], [9].

Another study demonstrates a malware detection framework which tracks and analyzes and detects threats which affect power supply and energy of the device. The frameworks main activities surround sampling of power consumption by the applications generating power signatures which are then used to detect malware by measuring how similar are these power signatures and using $\chi^2$ distance measure. Basic experiments applied to HP iPAQ device which has a windows mobile OS on it, showed 99% of classifying mobile malware [10].

Battery sensing intrusion detection system (B-SIPS) for mobile devices was developed by Buennemeyer et al. It deals with detecting abnormal current changes. B-SIPS relies on the SNORT IDS which provides a signature based detection of attack [11].

One study used linear regression model to estimate power consumption, and the model is based on parameters CPU-Load and disc access and its main objective is to determine the amount of energy used on a per-process basis and to allocate or find processes that could possibly exhaust the energy source [12].

Jacoby and Davis demonstrate IDS B-BID whose primary focus is detecting the attacks on battery source. This IDS monitors the devices electrical current and evaluates the correlation with the known signatures and patterns. In order to detect attacks the electrical current is measured and if the anomaly is detected all kinds of data such as network activity, CPU usage process count is collected and tested against known attack signatures. Additional feature is to send this data to the network administrator [13].

Miettinen presented a hybrid host-and-network based IDS, because of the claim that alone, these system are not sufficient, the engine on the server filters the received alarms according to correlation rules stored in the knowledge base and sends the results to the security monitoring GUI for the administrator to analyze [14].

Hwang evaluated the effectiveness of keystroke dynamics-based authentication (KDA) on mobile devices. Their empirical evaluation focuses on a short four digit pin numbers and the results show only 4% of false alert rate [15].

The most recent research on this topic is the KBTA framework[16]. A new kind of host-based IDS for smartphones is proposed in which detection is achieved by continual monitoring of mobile device and looking for malicious temporal behaviour. The framework itself relies on a lightweight agent in terms of CPU, memory and power consumption. This agent continuously samples different features on the device, performs the analysis of the gathered data and then infers the state of the device. The reason why this approach is unique is because of the use of temporal framework which utilizes the behavioural patterns registered over time. This is then used to discover previously unknown malware based on these patterns.

This framework also requires that the interpretation of the gathered data should be context sensitive and that it should be analyzed within given context rather then pure classification based on signature or methods that were used in previous work.

Oher major IDS proposed is Andromaly [17]. It is host-based and it continually monitors various features and events obtained from mobile device and then applies machine learning anomaly detectors to classify the collected data as normal-benign, or abnormal-malicious.

The classifiers that were examined include k-means, logistic regression, histograms, decision trees, Bayesian networks, naive Bayes. The evaluation is typically split into two phases – training and testing. In the first phase a training set of benign and malicious features is provided to the system in a form of feature vector. These features are both collected during normal operations of the system, reflecting benign behaviour and when malware is active on the system, thus providing vectors for malicious behaviours in order for the classifier to be properly trained.

One other major approach proposed is cloud-based IDS for smartphones[18]. It consists of a cloud-based service which would allow users to install a light-weight agent on their smartphones and register to an online cloud-service by specifying their operating system, applications installed on their phone and other relevant information about their device. Afterwards, this specific smartphone is emulated in a virtual machine on the cloud using a proxy which duplicates the incoming traffic to the device and then forwards the traffic to the emulation platform, where detection and forensic analysis is preformed.

## PROPOSED MODEL - AMOXID

This model proposes categorization of threats into three main categories: 1 – Threats to user's experience; 2 – Cost generating threats; 3 – Privacy infringing threats.

Threats to user's experience revolve around user's general experience and enjoyment while using the smartphone and they include:
  a) Malware that drains the battery will decrease the smartphone live-time and will necessitate often recharging which will become burdensome on the user.
  b) Malware which attacks and eats up memory will slow down the speed and responsiveness of the applications and the user experience will suffer.

Threats that generate cost include:
  a) Outgoing phone calls (within user's telephone network, in roaming mode, long distance calls)
  b) Outgoing SMS messages
  c) Unauthorized connection to GPRS internet while the user is in the roaming mode or while the user is in a foreign country, all of which can generate extremely large costs. Our model primarily focuses on this kind threats and extractions of feature that are related to generating of cost.
  d) Other cost that can be generated is in case of online shopping where user's credit card data is compromised, which is then used or sold on a black market causing financial damage to the user. This overlaps with privacy infringing threats.

Privacy infringing threats include the following:
  a) Logging of keystrokes.
  b) Recording user's voice conversations;
  c) Copying users documents such as images and text notes;
  d) Logging and copying SMS messages, phone call logs, user's phone book;
  e) Watching user's activity on the internet, such as which websites he is visiting, his accounts passwords and other Internet activity;
  f) Tracking users Geo-location and user movement (GPS);

The above categories can be analyzed separately and dealt with as three different subsystems in the IDS for smarphones.

*AmoxID - Network policies and conceptual model*

Smartphone is a device to be carried as user moves between different locations. When moving between different locations user enters different types of network. At home he uses his own wireless network within his household. At work user is within his company's network. On the outside user connects to outside networks. Different networks have different threats lurking around and some networks have Network IDS employed (such as company's network). Users of smartphones are sometimes given smartphones by their companies which require them to follow policies and which additionally need to protect their employee's phones, because not protecting their employees would mean not protecting companies data. We introduce the concept of setting different policies and IDS operating with different levels of alertness depending upon which network is user connected to.

This paper has primary goal of giving a conceptual model of possible IDS designed for mobile devices. The results achieved on the chosen vector parameters for the SVM could only be valid if tested on different smartphone devices running same operating system with different processing and memory power. Multi-core systems, GB-RAM memories are available only to an expensive palette of phones so the performance of the IDS on these systems would produce better results and higher accuracy of classification. All of these will be left for future work and will be tested in future papers and work on this issue.

The proof of concept IDS proposed and developed for Android is named AmoxID. The conceptual model is given in Figure 1.

This model proposes system of policies where depending on the user's current network, different policy level is applied and detection mechanism is on higher alert in public networks, while home and work networks are to be considered safer and there is a possibility of integration with Network IDS of the company or IDS that user sets up at his own household network.

This can be of importance to companies who are issuing phones to their employees. If company sends confidential emails and gives confidential data to employees that are accessing this through the smartphone, it is important that this information stays protected. Specially designed policies for employee's smartphones that are issued to employees can be configured to include pre-built IDS which would enforce different policies depending on the user's current network. If the user happens to be outside of company's network, different policies are applied which protect access to company's data on the phone. When inside the company's network, the installed IDS could be maintained by the network administrator. IDS would interact with IDS on the server such as SNORT which could then utilize higher processing and memory power to analyze logs on the employee's smartphone. All of this can also be designed in a way not to diminish users overal experience and without much effort on the user's side.

Other thing that should be taken into consideration is the fact that there is higher chance for power source of the user's smartphone being available when the user is within his home or work environment.

This paper proposes three types of policies to be enforced depending on te users current network.
1.Home Network, 2.Work Network, 3.Public Network (inside of user's telephone network provider or outside)

Features that could are tracked and based on which we create vectors for SVM in this version of AmoxID IDS are as follows:

1) 1.**call_ntinlst_inn**: Number of outgoing calls toward numbers outside contact list within user's telephone network provider.
2) 2.**call_ntinlst_outn**: Number of outgoing calls toward numbers outside contact list outside user's telephone network provider.
3) 3.**call_inlst_inn**: number of outgoing calls toward numbers that are inside of contact list inside user's telephone network provider.
4) 4.**call_inlst_outn**: Number of outgoing calls towards numbers that are inside of contact list outside user's telephone network provider.
5) 5.**sms_ntinlst_inn**: Number of SMS going towards numbers outside of contact list within user's telephone network provider.
6) 6.**sms_ntinlst_outn**: Number of SMS going towards numbers outside of contact list outside user's telephone network provider.
7) 7.**sms_inlst_inn**: Number of SMS going towards numbers inside of contact list within user's telephone network provider.

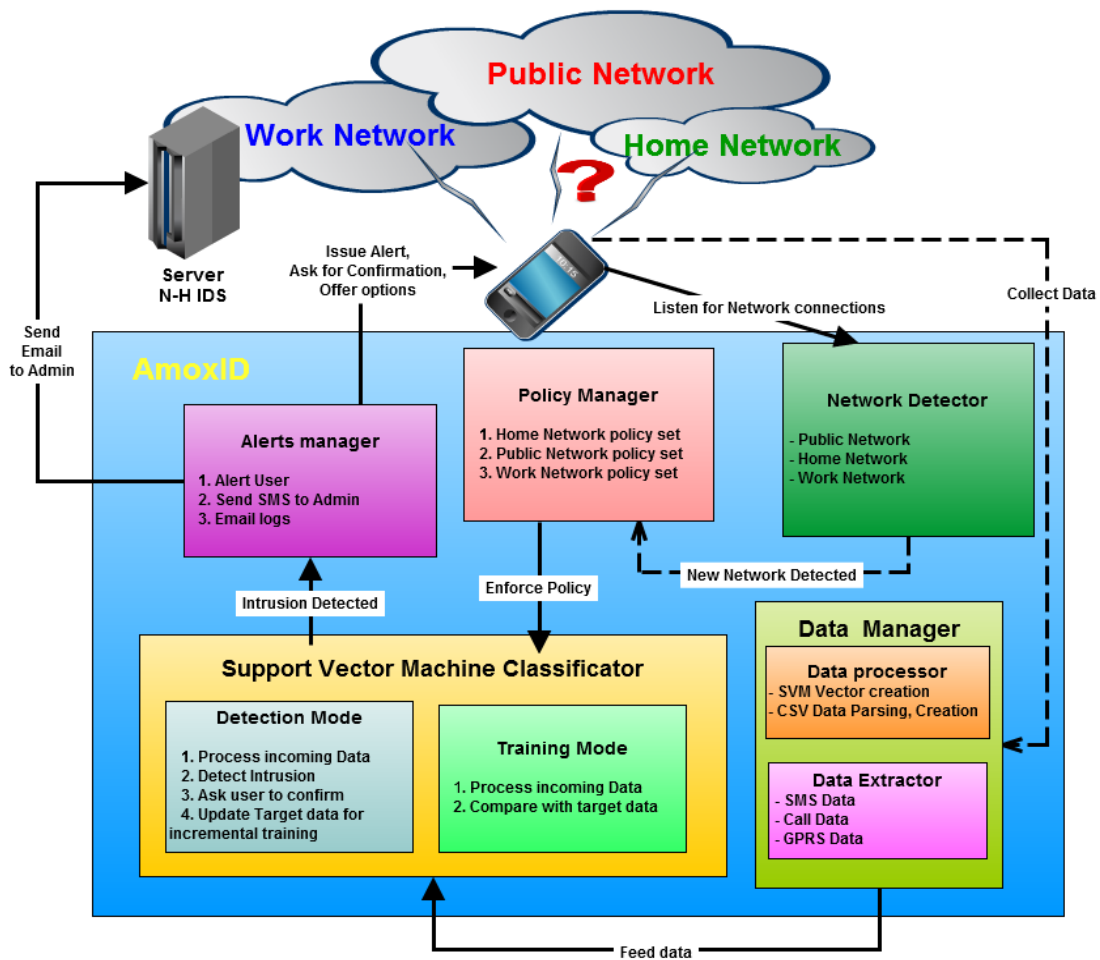

**Figure 1: AmoxID – Conceptual Model**

8) 8.**sms_inlst_outn**: Number of SMS going towards numbers inside of contact list outside user's telephone network provider.
9) 9.**gprs_inn**: Connection to GPRS within user's telephone network provider.
10) 10.**gprs_outn**: Connection to GPRS outside user's telephone network provider.

The screenshots of AmoxID application developed for this purpose are given below:

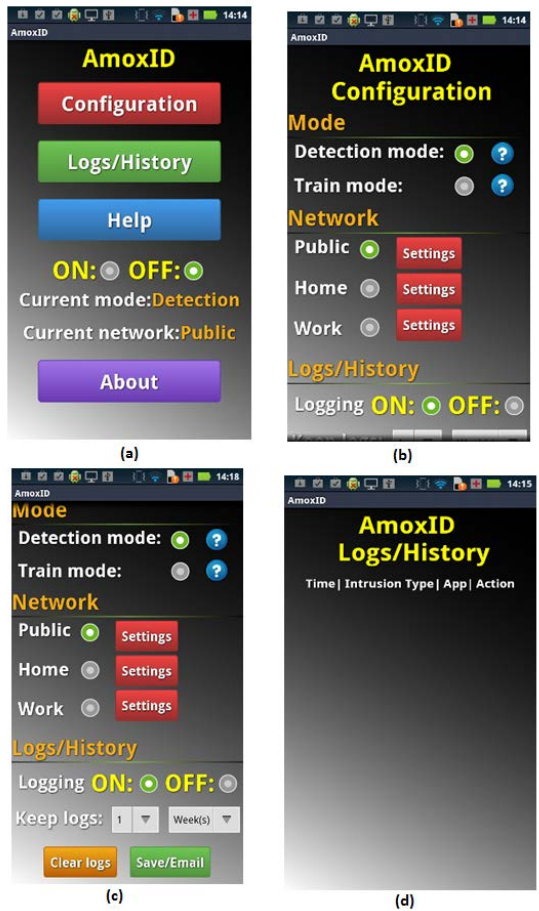

**Figure 2: AmoxID – Android Application**

## V. CONCLUSIONS AND FUTURE WORK

This paper has given a brief overview of Intrusion Detection Systems used in various contexts of computer devices security with special focus on smartphones. We have proposed a model based on SVM classification and proposed enforcement of different policies based on the type of Network the user happens to connect to. We have categorized threats on mobile devices into three major categories which can be addressed separately. These include 1 - user-experience threats, 2 - cost generating threats, 3 – privacy-infringing threats.

The future work will involve thorough testing of AmoxID on different devices with Android operating system and consequent results will be published in relevant academic literature.